\documentclass[aps,prb,superscriptaddress,twocolumn,showpacs]{revtex4-1}
\usepackage[utf8]{inputenc}
\usepackage[american,british]{babel}
\usepackage[T1]{fontenc}
\usepackage[pdftex]{graphicx}  
\usepackage{xcolor}
\usepackage{dcolumn}
\usepackage{bm}
\usepackage{amsmath,amsthm,amssymb,mathrsfs,mathtools,amsfonts,braket}
\usepackage{verbatim}
\usepackage{ulem}
\usepackage{epsfig}
\usepackage{color}
\usepackage{xfrac}
\usepackage{geometry}
\usepackage{hyperref}

\begin{document}

\title{Exact many-body scars and their stability in constrained quantum chains}

\author{Federica Maria Surace}
\affiliation{SISSA, via
Bonomea 265, 34136 Trieste, Italy}
\affiliation{The Abdus Salam International Center for Theoretical Physics, Strada Costiera 11, 34151 Trieste, Italy}

\author{Matteo Votto}
\affiliation{Dipartimento di Fisica, Università degli Studi di Milano, Via Celoria 16, 20133, Milano}
\affiliation{The Abdus Salam International Center for Theoretical Physics, Strada Costiera 11, 34151 Trieste, Italy}
\affiliation{SISSA, via Bonomea 265, 34136 Trieste, Italy}

\author{Eduardo Gonzalez Lazo}
\affiliation{SISSA, via
Bonomea 265, 34136 Trieste, Italy}
\affiliation{The Abdus Salam International Center for Theoretical Physics, Strada Costiera 11, 34151 Trieste, Italy}

\author{Alessandro Silva}
\affiliation{SISSA, via
Bonomea 265, 34136 Trieste, Italy}

\author{Marcello Dalmonte}
\affiliation{The Abdus Salam International Center for Theoretical Physics, Strada Costiera 11, 34151 Trieste, Italy}
\affiliation{SISSA, via
Bonomea 265, 34136 Trieste, Italy}

\author{Giuliano Giudici}
\affiliation{SISSA, via
Bonomea 265, 34136 Trieste, Italy}
\affiliation{The Abdus Salam International Center for Theoretical Physics, Strada Costiera 11, 34151 Trieste, Italy}

\date{\today}

\begin{abstract}
Quantum scars are non-thermal eigenstates characterized by low entanglement entropy, initially detected in systems subject to nearest-neighbor Rydberg blockade, the so called PXP model. While most of these special eigenstates elude an analytical description and seem to hybridize with nearby thermal eigenstates for large systems, some of them can be written as matrix product states (MPS) with size-independent bond dimension. We study the response of these exact quantum scars to perturbations by analysing the scaling of the fidelity susceptibility with system size. We find that some of them are anomalously stable at first order in perturbation theory, in sharp contrast to the eigenstate thermalization hypothesis. However, this stability seems to breakdown when all orders are taken into account. We further investigate models with larger blockade radius and find a novel set of exact quantum scars, that we write down analytically and compare with the PXP exact eigenstates. We show that they exhibit the same robustness against perturbations at first order.
\end{abstract}

\maketitle

\section{Introduction}
The eigenstate thermalization hypothesis (ETH)~\cite{Deutsch1991,Srednicki1994} legitimises the use of quantum statistical mechanics to describe the equilibrium properties of isolated many-body systems emerging from their coherent dynamics. In a nutshell, it states that the expectation values of physical observables on finite-energy density eigenstates of the Hamiltonian yield a smooth function of the energy for large systems, the off-diagonal matrix elements being pseudo-random numbers. The range of validity of this assumption encompasses a wide variety of interacting systems~\cite{}, but non-generic exceptions have been found. In fact, the presence of conservation laws is known to prevent thermalization in integrable systems, due to a breakdown of ETH~\cite{Rigol2008}. A similar scenario occurs in the presence of strong disorder, when energy eigenstates localize~\cite{mbl0,mbl1,mbl2,mbl3,mbl4}.

More recently, ETH violations have been detected in systems whose long-time steady state looks thermal for most of the initial states. Only when specific initial conditions are chosen, the dynamics is anomalously slow (when compared to the majority of other choices of initial states) and thermalization is not observed on experimentally accessible time-scales~\cite{Bernien2017}. The origin of this phenomenology is the presence in the energy spectrum of a few eigenstates, dubbed many-body quantum scars, that do not obey ETH and possess a large overlap with the initial state at hand. They are characterized by expectation values of local observables that do not agree with the canonical ensemble at their energy and by a sub-extensive entanglement entropy~\cite{Turner2018,Turner2018a,Khemani2019,Choi2019,Ho2019,Motrunich2019,Iadecola2019,Michailidis2019,Moudgalya2020,hart2020random}. 

The archetypal model in which these special eigenstates arise is the PXP model~\cite{fss2004,lesa2012}, introduced as a simplified description of the Rydberg atom chain realized in Ref.~\onlinecite{Bernien2017}. As a consequence of the effective interaction between Rydberg states, the experimental setup simulates a spin-1/2 system with a tunable parameter --the blockade radius $R_b$-- that describes how certain states, where two spins up are separated by less than $R_b$ lattice sites, are never explored by the dynamics due to a large energy penalty. In the simplified description $R_b$ becomes a discrete parameter which we will call $\alpha$ in what follows, where $\alpha=1$ is the PXP model.

Quantum scars in the PXP model were originally used to explain the slow dynamics observed by evolving a charge-density wave (CDW) initial state in the above-mentioned experiment with Rydberg atoms: for a chain of length $L$, there are $L+1$ scar eigenstates, with a large overlap with the CDW, spread throughout the spectrum and (approximately) equally spaced in energy. Crucially, numerical results reveal hybridization of these scars with thermal eigenstates, implying that they are not stable in the thermodynamic limit~\cite{Turner2018a}. Therefore 
the resulting dynamics from this initial state is expected to eventually thermalize. However, two exact uniform matrix product eigenstates have been found for all (even) system sizes~\cite{Motrunich2019}. This fact demonstrated the existence of ETH violating eigenstates that survive in the infinite size limit, and motivated the study of 
their stability against perturbation.
In Ref.~\onlinecite{Motrunich2020} the authors address this problem by using perturbation theory: from the scaling of the averaged matrix elements, they find no qualitative difference between the scars and thermal eigenstates, and thus deduce that the scars are not stable against perturbations. Nonetheless, they claim that thermalization is slow, because of parametrically small matrix elements.

Here, we analyse a different quantity (the fidelity susceptibility), which is a renowned probe of quantum chaos~\cite{Kuba1,Kuba2,Polkovnikov2020}, and is not subject to the arbitrariness of the averaging procedure. Part of our results contrast with Ref.~\onlinecite{Motrunich2020}, showing that the scars with zero energy have a completely different behavior from thermal eigenstates and are anomalously stable to first order in perturbation theory. These findings suggest that thermalization of quantum scars is even slower than previously expected, being originated from effects beyond the first perturbative order.

We remark that this anomalous stability is observed only for scars with zero energy, so we cannot conjecture a similar mechanism for explaining the persistence of non-exact scars at finite energy in the PXP model. 
In fact, although a construction based on a "single mode approximation" suggests a possible connection between 
the band of $L+1$ quantum scars at all energies to the MPS quantum scars at zero energy~\cite{Motrunich2019}, these two sets of low-entropy eigenstates appear to have different origin. For example, while the former are stabilized by a specific fine-tuned perturbation~\cite{Choi2019} and have logarithmic scaling of entanglement entropy with system size,
the latter are destroyed by the same perturbation and have finite entanglement entropy in the thermodynamic limit. 

In order to frame our finding about scar stability in the broader picture of ETH violations in constrained quantum systems, we prove that a novel set of exact eigenstates arising at zero energy (and at non-zero energy, when open boundary conditions are imposed) exists in generalized PXP models with $\alpha>1$. We do not find a band of eigenstates equally spaced in energy like the one observed in the PXP model. These results suggest that exact scars are a generic property of one-dimensional models constrained by Rydberg blockade. 
We then extend our stability analysis to this second set of scars, and show how, in analogy with the $\alpha=1$ case, they display anomalous stability. 

The paper is structured as follows. In Sec.~\ref{sec:PXP}, we introduce the PXP model and the scar eigenstates, and we set the notation for the following sections. In Sec.~\ref{sec:ETH} we introduce the fidelity susceptibility and the eigenstate thermalization hypothesis, and put forward a link between such observable and a recently proposed spectral version of the adiabatic gauge potential~\cite{PhysRevX.10.041017,Polkovnikov2020}. In Sec.~\ref{sec:PXPA} we focus on the models with radius of constraints $\alpha>1$: we discuss their properties in light of the ETH, we show that they obey Wigner-Dyson spectral statistics (Sec.~\ref{subsec:WD}); we describe the exact scars with $E=0$ as product states of "dimers" (Sec.~\ref{subsec:zero}), and the exact scars with $E\neq 0$ as matrix product states (Sec.~\ref{subsec:nonzero}); finally, we show that the exact scars with $E=0$ are anomalously stable against perturbations (Sec.~\ref{subsec:fidsus}).

\section{PXP model}
\label{sec:PXP}

The model we consider is the PXP model. {This model was first introduced in the context of constrained quantum models that can be directly related, in some parameter regimes, to exactly soluble classical statistical mechanics systems~\cite{fss2004}. In Ref.~\onlinecite{Lesanovsky2012}, it was shown how the same type of dynamics describes Rydberg excitations in an atomic chain in the regime of nearest-neighbour blockade. Each atom of the chain is modelled by a spin $1/2$: the state $\ket{0}$ corresponds to the ground state and the state $\ket{1}$ is an excited Rydberg state with high principal quantum number. A laser can couple the two states, inducing single-atom Rabi oscillations (in most experimental scenarios, such transition is actually driven by a pair of laser fields, via an intermediate, low-lying excited state). In the nearest-neighbour blockade regime, the interaction between Rydberg states on neighbouring sites is so large that the dynamics is effectively constrained to the subspace generated by the states with no consecutive "1"s.}

Defining $X_i,Y_i,Z_i$ as the Pauli matrices at site $i$ and $P_i=(1-Z_i)/2$, $n_i=(1+Z_i)/2$, the dynamics in the constrained space is described by 

\begin{equation}
    H_0=X_1P_2+\sum_{j=2}^{L-1}P_{j-1}X_jP_{j+1}+P_{L-1}X_L
\end{equation}
for open boundary conditions and

\begin{equation}
    H_0=\sum_{j=1}^{L}P_{j-1}X_jP_{j+1}
\end{equation}
with the identification of the sites $j\equiv j+L$ for periodic boundary conditions.
{Because of Rydberg blockade, the Hamiltonian acts on the space constrained by the conditions $n_i n_{i+1}=0$ for every $i$.}

We are interested in the effects induced by a perturbation $V$ that has the same symmetries of $H_0$. 
More concretely, the Hamiltonian is $H=H_0+\lambda V$, where

\begin{multline}
    V=X_1P_2Z_3+\sum_{j=2}^{L-2}P_{j-1}X_jP_{j+1}Z_{j+2}\\
    +\sum_{j=3}^{L-1}Z_{j-2}P_{j-1}X_jP_{j+1}+Z_{L-2}P_{L-1}X_L
\end{multline}
for the case of open boundary conditions and

\begin{equation}
    V=\sum_{j=1}^{L}(P_{j-1}X_jP_{j+1}Z_{j+2}+Z_{j-2}P_{j-1}X_jP_{j+1})
\end{equation}
for periodic boundary conditions.

Both $H_0$ and $V$ commute with the space reflection symmetry $I$ and anticommute with the particle-hole symmetry $C_{ph}=\prod_i \sigma_i^z$. As a consequence, the spectrum is symmetric with respect to the eigenvalue $E=0$ and the energy zero eigenspace has a dimension growing exponentially with system size~\cite{Schecter2018}. For more details about the peculiar properties of the spectrum we refer to Appendix~\ref{sec:prop}.

\subsubsection{Many-body scars}
As stated above, many-body scars are states that do not satisfy ETH. {It was shown in Ref.~\onlinecite{Turner2018} that the spectrum of the PXP model exhibits a band of equally-spaced many-body scars. These scars were responsible for the observation of long-lived oscillation in a Rydberg atom experiment~\cite{Bernien2017}. Their exact form is not known analytically, and their persistence in the thermodynamic limit is still an open question. However, as was shown in Ref.~\onlinecite{Motrunich2019}, $H_0$ has also some exact scars in the form of MPS eigenstates at finite energy density.} For open boundary conditions they are defined as
\begin{equation}
\label{eq:scarsobc}
    \ket{\Gamma_{i,j}}=\sum_{\{\sigma\}}v^T_i A_{\sigma_1\sigma_2}\dots A_{\sigma_{L-1}\sigma_{L}}v_j \ket{\sigma_1\sigma_2\dots\sigma_{L-1}\sigma_L}
\end{equation}
with
\begin{equation}
    A_{00}=\begin{pmatrix}
    0 & -1\\
    1 & 0
    \end{pmatrix},\; A_{01}=\begin{pmatrix}
    \sqrt{2} & 0\\
    0 & 0 
    \end{pmatrix},\;
    A_{10}=\begin{pmatrix}
    0 & 0\\
    0 & -\sqrt{2} 
    \end{pmatrix},
\end{equation}
$i,j=1,2$ and $v_1=(1,1)^T$, $v_2=(1,-1)^T$. $\Gamma_{12}$ has energy $\sqrt{2}$, $\Gamma_{21}$ has energy $-\sqrt{2}$, whereas $\Gamma_{11}$ and $\Gamma_{22}$ have energy $0$. In the next sections, we will focus on scars with well-defined inversion quantum number, so we define $\ket{\Gamma_I}=(\ket{\Gamma_{11}}-\ket{\Gamma_{22}})/\sqrt{2-2\braket{\Gamma_{11}|\Gamma_{22}}}$.

For periodic boundary conditions, the two scarred eigenstates $\ket{\Phi_1}$ and $\ket{\Phi_2}$ are defined as 

\begin{equation}
\label{eq:scarspbc}
    \ket{\Phi_1}=\sum_{\{\sigma\}}\text{Tr}[A_{\sigma_1\sigma_2}\dots A_{\sigma_{L-1}\sigma_{L}}]\ket{\sigma_1\sigma_2\dots\sigma_{L-1}\sigma_L}
\end{equation}
and $\ket{\Phi_2}=T_x \ket{\Phi_1}$, {where $T_x$ is the translation operator}. Both have energy 0. Their properties under the symmetries are the following:
$I\ket{\Phi_i}=(-1)^{L/2}\ket{\Phi_i}$ and $C_{ph}\ket{\Phi_i}=(-1)^{L/2}\ket{\Phi_i}$ for $i=1,2$. We will work with the linear combinations
$\ket{\Phi_{K=0,\pi}}=(\ket{\Phi_1}\pm\ket{\Phi_2})/\sqrt{2\pm 2\braket{\Phi_1|\Phi_2}}$. Even though these are not responsible for the persistent oscillations observed in experiments, their putative stability in the thermodynamic limit outlines their importance.

\section{Perturbation theory and ETH}
\label{sec:ETH}
It is crucial to understand how to define {\it stability} for these kind of eigenstates. 
In general, we will say that an eigenstate of $H_0$ is stable if it can be deformed to an eigenstate of $H_0+\lambda V$ with a local unitary transformation in the thermodynamic limit. Usually this criterion is satisfied by ground states in gapped systems. Here we are interested in  
the scars $\ket{\Gamma_{\alpha\beta}}$ and $\ket{\Phi_i}$ which are in the middle of a dense spectrum, in the absence of a gap to protect them.
The local character of the transformation, if it exists, should guarantee that a stable scar retains its character (no ETH and area law entanglement) in the thermodynamic limit.
For generic eigenstates, no stability is expected. This can be understood as a consequence of the Eigenstate Thermalization Hypothesis (ETH): to first order in the perturbation strength $\lambda$, the perturbed eigenstate can be written as
\begin{equation}
    \ket{n^0}+\lambda\ket{n^1}=\ket{n^0}+\lambda \sum_{m\neq n}\frac{\braket{m^0|V|n^0}}{E_n^0-E_m^0}\ket{m^0}.
\end{equation}
According to ETH, the off-diagonal matrix element $\braket{m^0|V|n^0}$ scales as $\exp (-S/2)$, where $S$ is the extensive thermodynamic entropy of the system. The energy denominator, on the other hand, scales as $\exp(-S)$ for nearby eigenstates. This simple argument implies that the first order correction diverges exponentially in the system size $L$.

\begin{figure*}[t]
    \centering
    \includegraphics[width=0.95\textwidth]{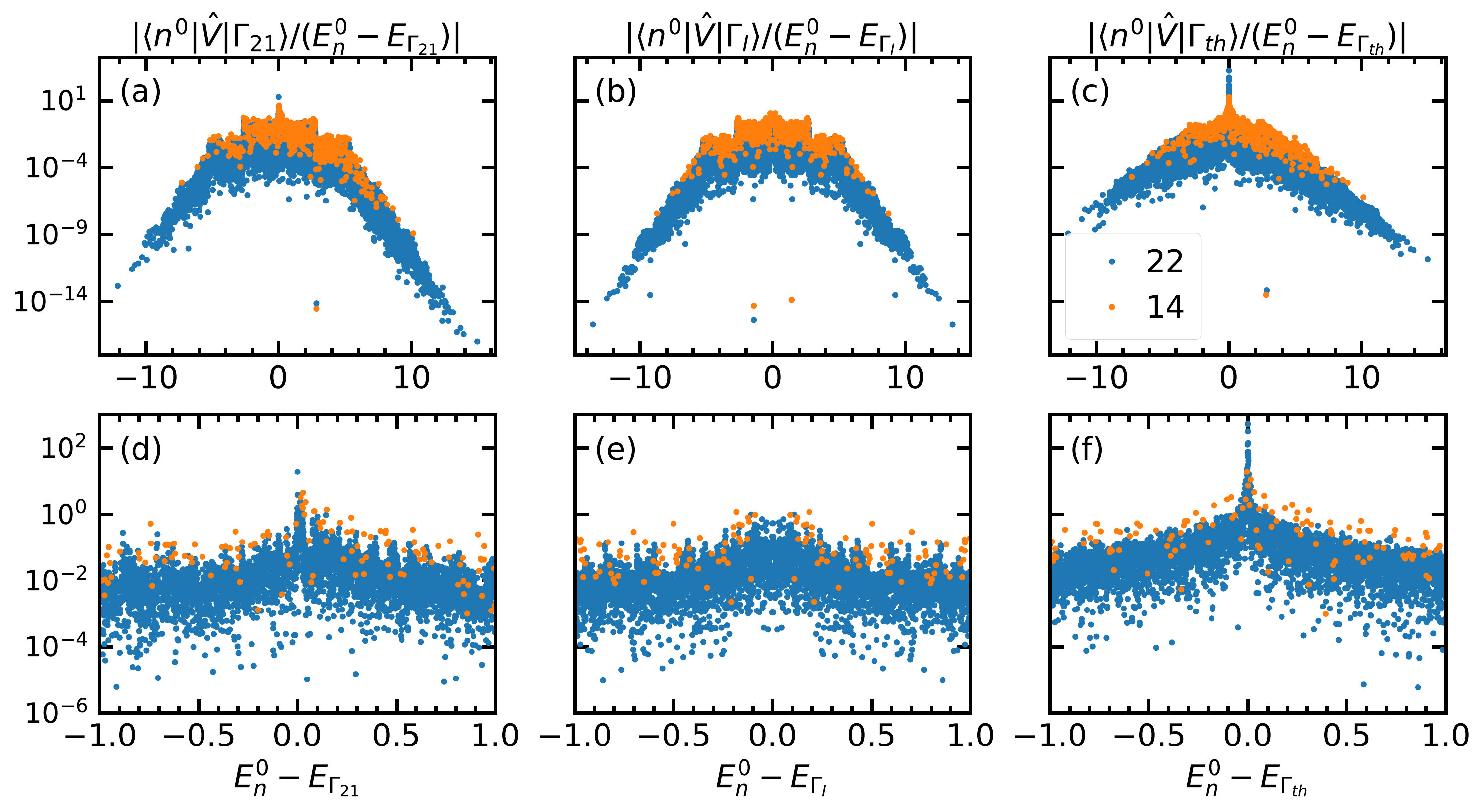}
    \caption{Absolute value of the ratio between the matrix element and the energy difference between a target state and a state of the spectrum. The same data are represented in a larger (first row) and in a smaller (second row) range of energy difference. The target states are the scars eigenstates $\ket{\Gamma_{21}}$ (a,d), $\ket{\Gamma_{I}}$ (b,e) defined in Sec.~\ref{sec:PXP} and a thermal eigenstate $\ket{\Gamma_{th}}$ (c,f) taken as the third eigenstate after $\ket{\Gamma_{21}}$ in order of increasing energy.
    The clear peak observed when a thermal eigenstate $\ket{\Gamma_{th}}$ is considered is not present for the scars eigenstates, pointing at a suppression of the matrix elements of the perturbation for the latter states.}
    \label{fig:matrixelements}
\end{figure*}

Hence, a natural question to answer is whether the first order correction to the scars behaves according to the scaling predicted by ETH or not. In Ref. \onlinecite{Motrunich2020}, it was found that the matrix elements $\braket{m^0|V|\Gamma}$ averaged over a certain set of eigenstates $\{\ket{m^0}\}$ close in energy to $\ket{\Gamma}$ do indeed scale as $\exp (-S/2)$, where $\ket{\Gamma}$ is one of the scars for the case of open boundary conditions. This is however not sufficient to claim instability: the matrix elements which are responsible for the divergence are the ones involving states that are very close in energy. As can be seen in Fig.~\ref{fig:matrixelements}, the matrix elements weighted with the inverse energy gaps behave very differently for the scars and for generic thermal states: the vanishing denominator produces a peak in the case of a thermal state; the scars, despite the vanishing energy gaps, do not exhibit this peak, signalling a suppression of matrix elements for small gaps. Moreover, the averaging procedure of matrix elements introduces some arbitrariness in this respect: the result depends on the choice of the set of eigenstates that are included in the average. 

In this work, we propose to diagnose the stability of scar eigenstates by studying the fidelity susceptibility, defined as~\cite{You2007}
\begin{equation}
    \chi_F\left[\ket{n^0}\right]=\lim_{\lambda \rightarrow 0} \frac{-2 \ln |\braket{n^0|n^\lambda}|}{\lambda^2}
\end{equation}
where $\ket{n^0}$ is an eigenstate of $H_0$ and $\ket{n^\lambda}$ is the eigenstate of $H_0+\lambda V$ obtained from $\ket{n^0}$ with a perturbative construction in $\lambda$. From the explicit construction of the state, one finds\footnote{We use that $\ket{n^\lambda}=(\ket{n^0}+\ket{n_\perp})/\lVert\ket{n^0}+\ket{n_\perp}\rVert$, with $\braket{n_\perp|n^0}=0$ to obtain $\braket{n^0|n^\lambda}=\lVert\ket{n^0}+\ket{n_\perp}\rVert^{-1}=(1+\braket{n_\perp|n_{\perp}})^{-1/2}=1-\frac{1}{2}\lambda^2\braket{n^1|n^1}+O(\lambda^3)$.}
\begin{equation}
    \chi_F\left[\ket{n^0}\right] =\sum_{m\neq n}\left\lvert\frac{\braket{m^0|V|n^0}}{E_n^0-E_m^0}\right\lvert^2.
\end{equation}
The fidelity susceptibility is a measure of the response of an eigenstate to perturbations: when averaged over different eigenstates, for example, it has been very recently used as a measure of quantum chaos~\cite{PhysRevX.10.041017, Polkovnikov2020}.
For gapped ground states of local Hamiltonians, it is expected to scale as $\chi_F\sim L$ with the system size $L$. On the other hand, as argued above, ETH implies a scaling $\chi_F \sim \exp (L)$ for 
eigenstates at finite energy density.

Note that, due to the special properties of this perturbation, all the matrix elements of $V$ between zero energy states vanish (see Appendix~\ref{sec:prop}): as a consequence, the fidelity susceptibility is well-defined even for states in the exponentially degenerate zero-energy manifold and can be computed for all the scarred eigenstates.

\begin{figure}
    \centering
    \includegraphics[width=0.492\textwidth]{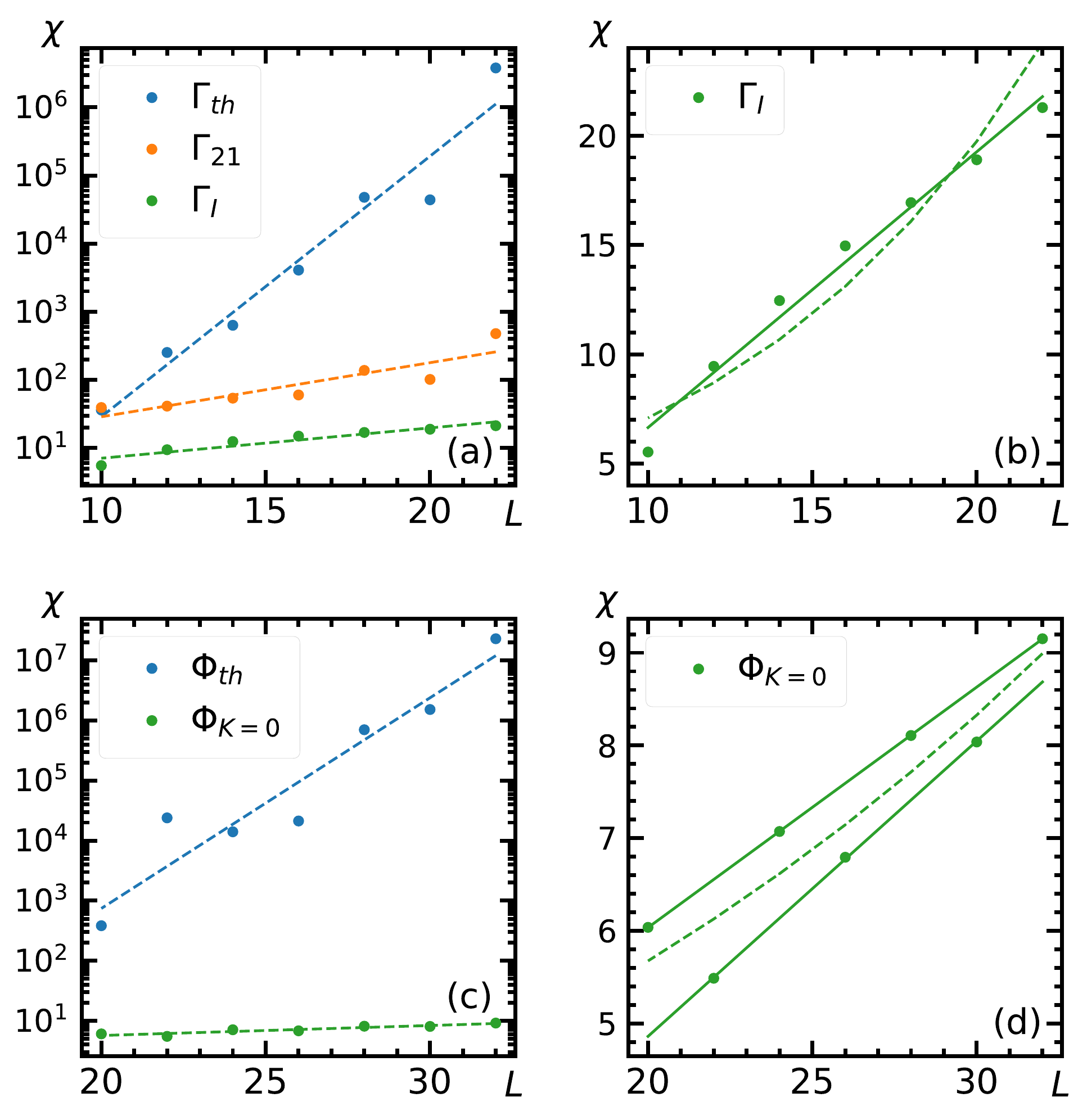}
    \caption{Scaling of the fidelity susceptibility with system size. The results shown refer to the states (a) $\ket{\Gamma_{th}}$, $\ket{\Gamma_{21}}$ and $\ket{\Gamma_{I}}$ with open boundary conditions and to the states (c) $\ket{\Phi_{th}}$ and $\ket{\Phi_{K=0}}$ with periodic boundary conditions. As can be seen in the panels with linear $y$-scale (b), (d), the scaling of the fidelity susceptibility  of a zero energy scar eigenstate is  polynomial with the system size, in sharp contrast to what happens for thermal eigenstates or scars at non-zero energy (a),(c). Solid lines are fits for the linear scalings (the two different lines for $\Phi_{K=0}$ capture the even-odd effect), dashed lines are fits for exponential scalings.
    }
    \label{fig:fidsusc}
\end{figure}

We obtain that only a subset of the exact scars appear to be stable. Indeed the scaling of the fidelity susceptibility for the scars $\ket{\Phi_{K=0}}$ (for the case of periodic boundary conditions) and $\ket{\Gamma_{I}}$ (for the case of open boundary conditions) shown in Fig.~\ref{fig:fidsusc} suggests a linear dependence\footnote{On top of the linear growth, the scaling for the scar $\ket{\Phi_{K=0}}$ is subject to an even-odd effect related to the different parity under inversion symmetry of the state ($I=(-1)^{L/2}$).} $\chi_F \sim L$, as evidenced by the solid lines. On the contrary, the scaling for $\ket{\Gamma_{21}}$ and for the generic thermal eigenstates $\ket{\Gamma_{th}}$ and $\ket{\Phi_{th}}$\footnote{The state $\ket{\Gamma_{th}}$ is chosen as the third eigenstate after $\ket{\Gamma_{21}}$ in increasing order of energy. The state $\ket{\Phi_{th}}$ is the state with energy closest to $-0.3$.} are compatible with an exponential growth (dashed lines), as predicted by ETH. These results show that $\ket{\Phi_{K=0}}$ and $\ket{\Gamma_{I}}$ are perturbatively stable to an infinitesimal perturbation. We note that these differences are not only qualitatively manifest (power versus exponential scaling), but also quantitatively striking, so that the different scaling regimes can be diagnosed despite the fact that our analysis is limited to modest system sizes up to $L=32$ spins.

We now want to understand if they are also stable to a finite strength $\lambda$ of the perturbation. If these states were akin to gapped ground states, we would have expected stability to hold in the thermodynamic limit for a finite $\lambda$ as long as it is much smaller than the gap. The absence of a gap makes the quest for an energy scale associated with scars much less obvious.

\begin{figure}[h]
    \centering
    \includegraphics[width=0.49\textwidth]{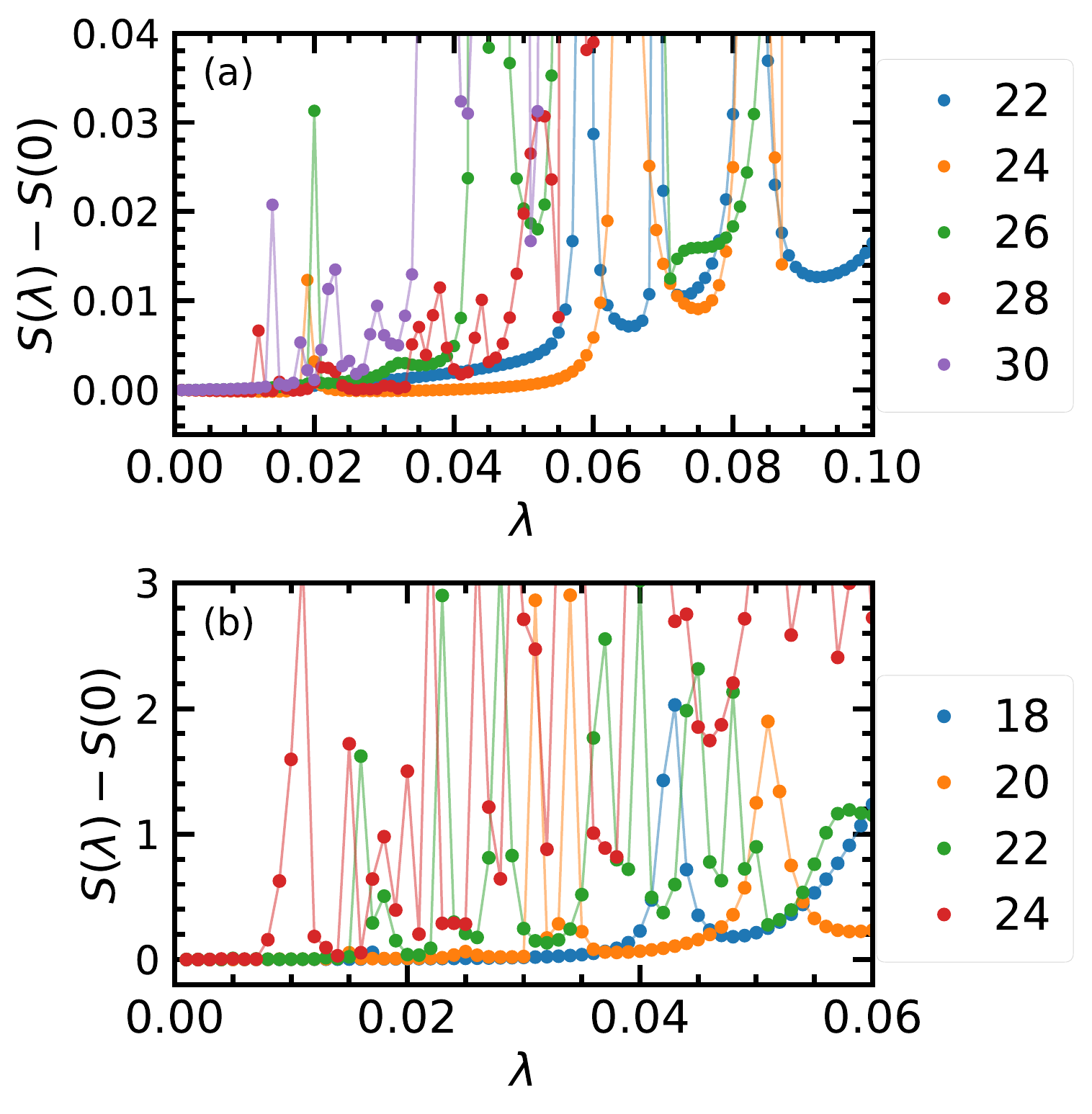}
    \caption{Bipartite entanglement entropy of the states (a) $\ket{\Phi_{K=0}^\lambda}$ and (b) $\ket{\Gamma_I^\lambda}$ as a function of $\lambda$. Peaks in this quantity signal hybridization of the perturbed state with thermal eigenstates. By increasing the system size, we find peaks closer and closer to $\lambda = 0$, suggesting that the scar eigenstates are not stable in the thermodynamic limit. }
    \label{fig:entropy}
\end{figure}

To address this problem, we compute the states $\ket{\Phi_{K=0}^\lambda}$ and  $\ket{\Gamma_i^\lambda}$ obtained by perturbing the scars $\ket{\Phi_{K=0}}$ and  $\ket{\Gamma_I}$ in the following way
\begin{equation}
    \ket{\Phi_{K=0}^\lambda}=\frac{1}{\mathcal N_\Phi^\lambda}\frac{1}{1+\lambda Q H_0^{-1}Q V}\ket{\Phi_{K=0}}
\end{equation}
\begin{equation}
    \ket{\Gamma_{I}^\lambda}=\frac{1}{\mathcal N_\Gamma^\lambda}\frac{1}{1+\lambda Q H_0^{-1}Q V}\ket{\Gamma_{I}}
\end{equation}
where $Q$ projects on the subspace with $E_0\neq 0$, and $\mathcal N_\Phi^\lambda$, $\mathcal N_\Gamma^\lambda$ are normalizing factors. The states $\ket{\Phi_{K=0}^\lambda}$ and $\ket{\Gamma_{I}^\lambda}$ are the perturbed eigenstates to infinite order in perturbation theory. We numerically compute the von Neumann bipartite entanglement entropy $S(\lambda)$ of these states for different system sizes (Fig.~\ref{fig:entropy}).
This quantity exhibits peaks that get closer to $\lambda=0$ as $L$ increases, indicating a stronger and stronger hybridization with other eigenstates in the spectrum. This fact strongly suggests that, despite the stability observed to first order in perturbation theory, the scars are ultimately not stable for finite $\lambda \neq 0$~\footnote{We note that performing a rigorous finite-size scaling analysis for the position of the first peak versus system size is tricky for two reasons: {\it (i)} we can only consider a coarse grained set of values of $\lambda$, so that we can only put an upper bound on the position of the peak, and {\it (ii)} the peaks may be due in principle to different level crossing, making a finite-size extrapolation not fully reliable. Our conclusion is based on the fact that we systematically observe the peak moving towards vanishing perturbations, with no exception, very rapidly with system size.}. 

\section{Models with radius of constraint $\alpha>1$}
\label{sec:PXPA}

Since the first studies on the PXP model, several other instances of quantum many-body scars have been put forward~\cite{Moudgalya2018,Moudgalya2018a,Mark2019,Schecter2019,Bull2019,Ok2019,Hudomal2019,Shibata2019,Chattopadhyay2019,Pai2019,Moudgalya2019,Surace2019,Surace2020,Iadecola2020,Mark2020,Zhao2020,Lee2020}. While it is tempting to extend some of the findings above to a general setting, we refrain from this for the very simple reason that PXP models have a characteristic feature - a constrainted Hilbert space that cannot be reduced in tensor product form - that is not present in other instances of quantum scars. We pursue instead an alternative route, based on investigating the stability of quantum scars in an enlarged class of constrained models.

In concrete, we consider a generalization of the PXP model, where we extend the constraint to the sites within an integer radius $\alpha$, i.e. $n_i n_j=0$ whenever $|i-j|\le \alpha$, with $n_j = \frac{Z_j+1}{2}$. The Hamiltonian has the form:

\begin{equation}
    H_0^\alpha=\sum_i P_{i-\alpha}\dots P_{i-1}X_iP_{i+1}\dots P_{i+\alpha},
    \label{eq:Halpha}
\end{equation}
where $P_j$ is the projector on the state $\ket{0}$. The Hamiltonian (\ref{eq:Halpha}) coincides with the PXP model for $\alpha=1$ and arises as an effective approximation of the long-range Hamiltonian describing Rydberg atoms arrays when the (continuous) blockade radius is increased (by e.g. tuning the distance between the atoms). Similarly to the PXP model, this Hamiltonian commutes with the reflection symmetry $I$ and anticommutes with the particle-hole symmetry $C_{ph}$, and the spectrum has the same properties (see Appendix~\ref{sec:prop}). 

\subsection{Spectral statistics}
\label{subsec:WD}
\begin{figure}[h]
    \centering
    \includegraphics[width=0.38\textwidth]{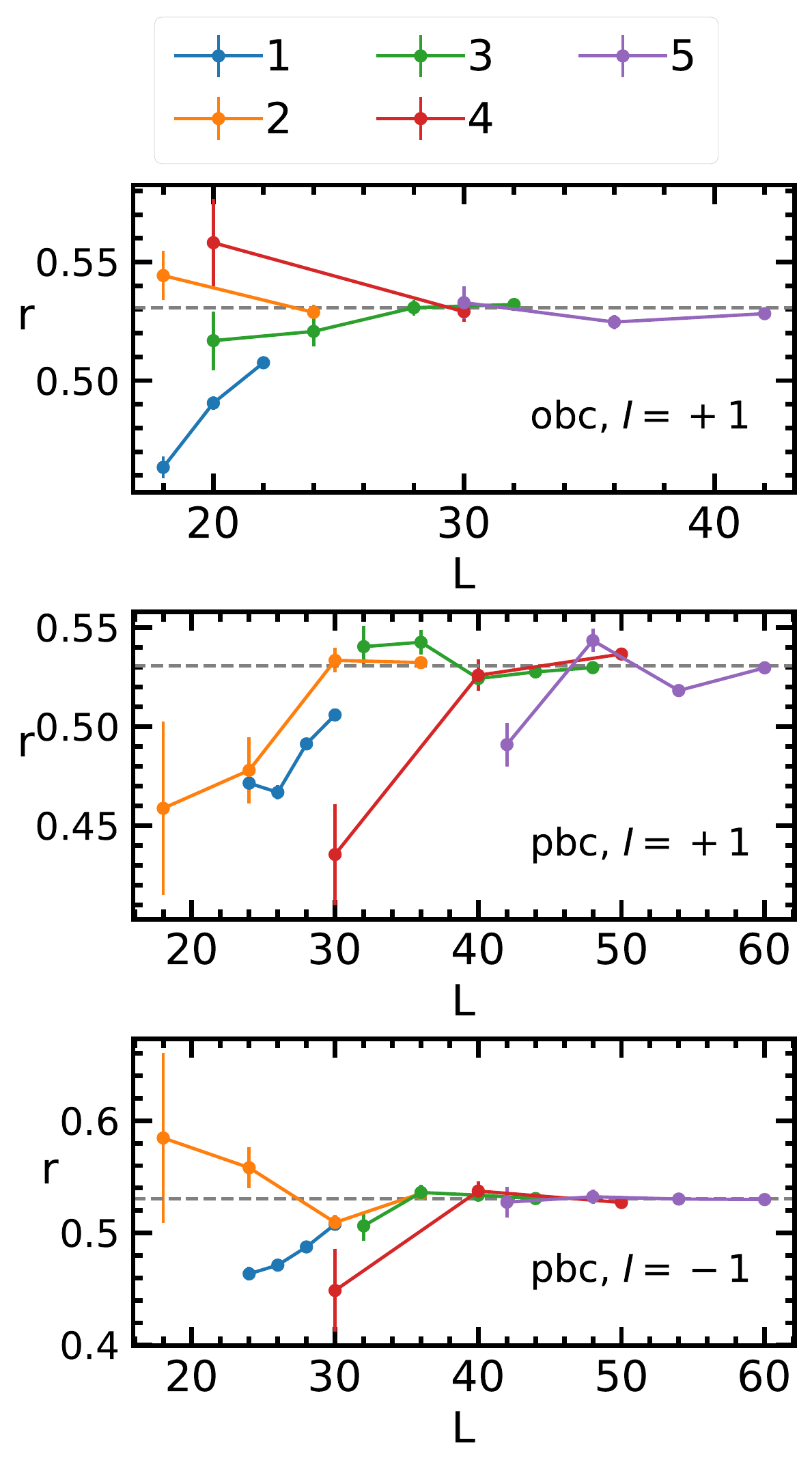}
    \caption{Ratio $r$ of nearby gaps averaged over the full spectrum. The colors label different values of $\alpha$. The dashed horizontal line is the value $r_{WD}$ associated with Wigner-Dyson statistics, that appears to be satisfied for all the values of $\alpha$ considered. }
    \label{fig:lsr}
\end{figure}

In this section we analyze the spectral statistics of the Hamiltonian in Eq.~(\ref{eq:Halpha}) for different values of $\alpha$. We use as a measure the ratio between nearby gaps:
\begin{equation}
r= \Big\langle \frac{ \mathrm{Min} \{ \Delta E_n  , \Delta E_{n + 1 } \}  }{ \mathrm{Max} \{ \Delta E_n  , \Delta E_{n + 1 } \} }\Big\rangle,
\label{r_ratio}
\end{equation}
where the average is taken over the full spectrum. For an ergodic system, this quantity is expected to flow to the value $r_{WD}\simeq 0.53$ associated with a Wigner-Dyson statistics. While for $\alpha=1$ ergodicity has been already verified in various works~\cite{Turner2018a,Khemani2019}, we check this assumption when $\alpha>1$ in Figs.~\ref{fig:lsr}, where we show the values of $r$ for different $\alpha$ and system sizes. In all the cases considered (reflection sector $I=+1$ with open boundary condition, reflection sectors $I=+1$ and $I=-1$ with momentum $K=0$ and periodic boundary conditions) we find a clear flow to $r_{WD}$ for increasing system sizes. We can therefore argue that the system has a spectral statistics that is compatible with ergodicity.

\subsection{Exact scars with $E=0$}
\label{subsec:zero}
We now show that, although the models considered here satisfy the Wigner-Dyson spectral statistics, some states in the spectrum have finite entanglement entropy in the thermodynamic limit and hence violate the eigenstate thermalization hypothesis. 

For a system with $L=(\alpha+2)n$ (with $n$ integer), consider the following state

\begin{equation}
\label{eq:scarphi}
\ket{\phi_\alpha}=\bigotimes_{i=0}^{n-1} \big[(\ket{01}-\ket{10})\ket{\underbrace{0\dots 0}_{\alpha}}\big]_{b_i}
\end{equation}
where the index $b_i$ labels blocks of $\alpha+2$ sites. The state of the first two sites of a block is an antisymmetric superposition (that we call a {\it dimer}) and hence is annihilated by the spin flip. All the other sites of a block cannot be flipped: they are "frozen" by the previous or the next dimer. Therefore, the state $\ket{\phi_\alpha}$ (and all the states obtained from it by translations) is a scar with energy $E=0$ for generic $\alpha>1$.

\begin{figure}[h]
    \centering
    \includegraphics[width=0.49\textwidth]{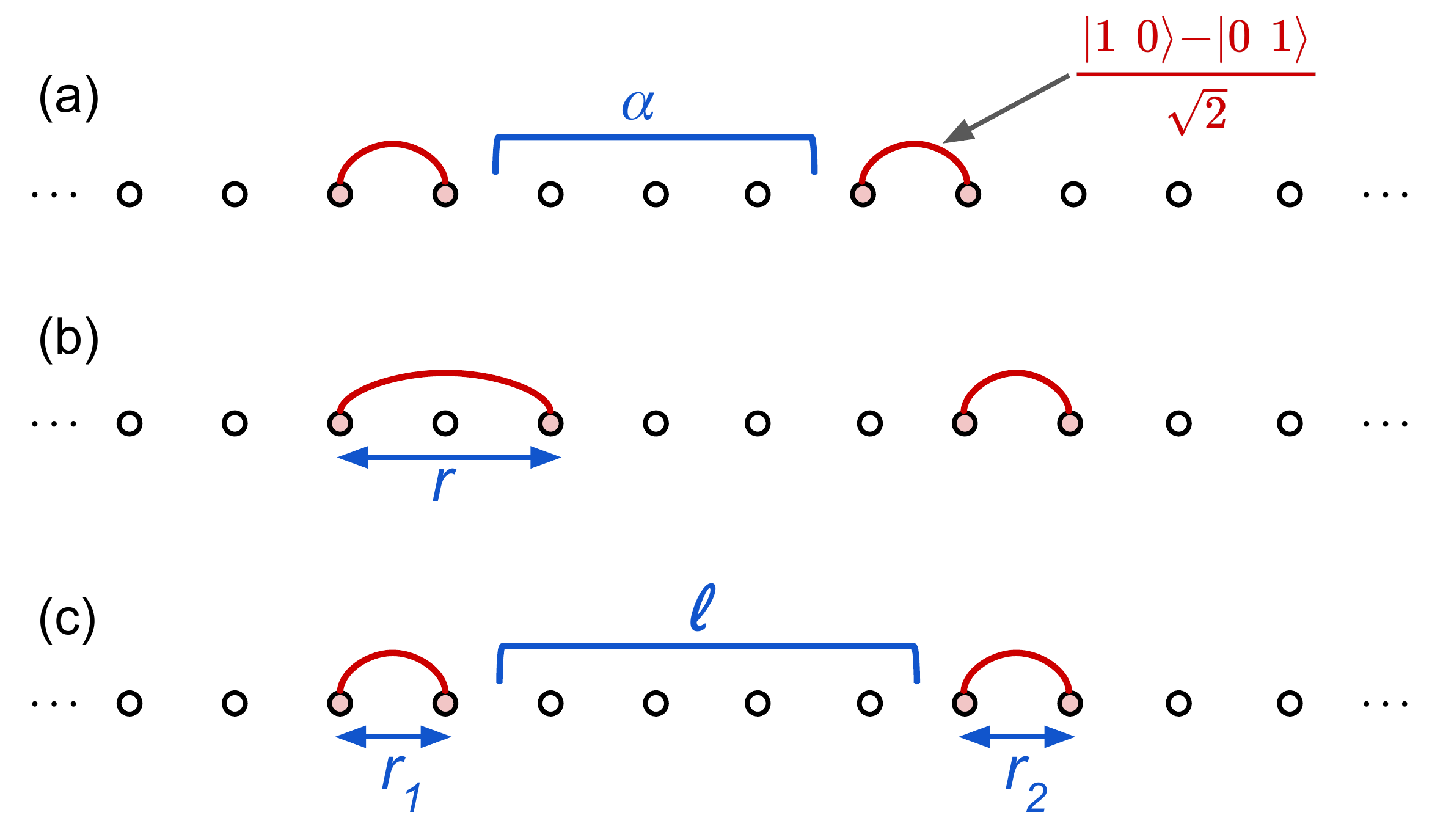}
    \caption{Some eigenstates  with $E=0$. (a) The state $\ket{\phi_\alpha}$ is made of dimers (in red) separated by sequences of $0$s of length $\alpha$. (b) Dimers can have range $r>1$. (c) Dimers can be separated by any distance $\ell$, such that $\alpha \le \ell \le 2\alpha-r_1-r_2$.}
    \label{fig:scars}
\end{figure}

We can construct many exact scars with $E=0$ by placing dimers (depicted in red in Fig.~\ref{fig:scars}) on the chain. Two dimers must be separated by a number $\ell$ of zeros in the range $\alpha \le \ell \le 2\alpha-2$. We can also have longer-range dimers involving sites that are not nearest neighbours. In this case, the number $\ell$ of zeros between two dimers of range $r_1$ and $r_2$ must be in the interval $\alpha \le \ell \le 2\alpha-r_1-r_2$. This last condition implies that the ranges of two consecutive dimers are bounded by $r_1+r_2 \le \alpha$.

This construction works also in the case of open boundary conditions, with the following rules for the boundaries: if the first (last) dimer of the chain has range $r$, then the number of zeros preceding (following) it must be $\ell\le \alpha-r$. 

We note that the structure of these states, that we write as product states of dimers, is reminiscent of the construction of scar eigenstates found in other constrained models\cite{lin2020quantum,Surace2020}.

\subsection{Exact scars with $E\neq 0$}
\label{subsec:nonzero}
In the following, we will show that the models of Eq.~(\ref{eq:Halpha}) have scars also at $E\neq 0$ when open boundary conditions are imposed. While, as we have shown in Sec.~\ref{subsec:zero}, it is possible to write many exact $E=0$ eigenstates as product states of dimers, for these scars we need to resort to a more involved construction: we write them as matrix product states with finite bond dimension, independent of the system size. 

\subsubsection{Exact scars with $E=\pm \sqrt{3}$}
\label{subsubsec:sqrt3}
For system sizes $L=(\alpha+2)n+3$, with $n$ integer, we are able to write two exact scars with energy $E=\pm \sqrt{3}$ as matrix product states. To define these states, we divide the chain in blocks labelled from $1$ to $2n+1$: the blocks labelled by odd numbers contain 3 sites, while the blocks labelled by even number contain $\alpha-1$ sites. As we prove in Appendix \ref{sec:firstproof}, the following state is an exact eigenstate with energy $E=\sqrt{3}$:

\begin{multline}
\label{eq:s3}
    \ket{\psi_\alpha^{(3)}}=\sum_{\vec s }\Big[ (1,0)^T \cdot N^{s_1}M^{s_2}\dots\\
    \dots M^{s_{2n}}N^{s_{2n+1}}\cdot (0,1) \Big] \ket{\vec s}
\end{multline}
where $s_1, s_2, \dots, s_{2n+1}$ label the states of the blocks and

\begin{equation}
    M^s = \begin{cases}
    1 & \text{if } s=00\dots00\\
    0 & \text{otherwise},
    \end{cases}
\end{equation}

\begin{equation}
\label{eq:N1}
N^{000}=\begin{pmatrix}
0 & \sqrt{3}\\
0 & 0\end{pmatrix},
\qquad N^{100}=\begin{pmatrix}
0 & 1\\
0 & 1\end{pmatrix},
\end{equation}
\begin{equation}
\label{eq:N2}
N^{010}=\begin{pmatrix}
1 & 1\\
0 & -1\end{pmatrix},\qquad
N^{001}=\begin{pmatrix}
-1 & 1\\
0 & 0\end{pmatrix}.
\end{equation}
 
From the relation $C_{ph}H_0^\alpha=-H_0^\alpha C_{ph}$ we immediately find that the state $\ket{\psi_\alpha^{(-3)}}=C_{ph}\ket{\psi_\alpha^{(3)}}$ is another eigenstate of $H_0^\alpha$ with eigenvalue $E=-\sqrt{3}$.\\
We also note that the state obtained by taking the trace in Eq.~\eqref{eq:s3} is a zero energy eigenstate for $L=(\alpha+2)n+3$ when open boundary conditions are imposed. Moreover, removing the matrix $N$ at one of the two boundaries we can construct an MPS that is invariant under translations of $\alpha+2$ sites
\begin{equation}
\label{eq:cat}
\ket{\varphi_\alpha} = \sum_{\vec{s}} \mathrm{Tr} ( B^{s_1} B^{s_2} \dots B^{s_n} ) \ket{\vec{s}},
\end{equation}
where $B = M N$ and $s_i$ runs through the $3$ allowed states of the $i$-th block, made of $\alpha+2$ sites.  
This state is a zero energy eigenstate for periodic boundary conditions and system sizes $L=(\alpha+2)n$, and it has non-vanishing overlap with the dimer eigenstates of Sec.~\ref{subsec:zero}; however, for generic $\alpha$ it has a component that is independent of those states. The matrix $B$ yields a non-injective MPS, whose parent Hamiltonian has a degenerate groundspace~\cite{pgarcia2008}. In fact, the state in Eq.~\eqref{eq:cat} can be written as a cat state
\begin{align}
\nonumber
\ket{\varphi_\alpha} & = \left[ \left( \ket{L} + \frac{1}{2}\ket{R} -\frac{3}{2}\ket{C} \right) \ket{\underbrace{0\dots 0}_{\alpha-1}} \right]^{\otimes n }  \\[1mm]
\nonumber
& +  \left[ \left( \frac{1}{2} \ket{L} + \ket{R} -\frac{3}{2}\ket{C} \right) \ket{\underbrace{0\dots 0}_{\alpha-1}} \right]^{\otimes n } \\[1mm]
& = \ket{\varphi^1_\alpha} + \ket{\varphi^2_\alpha},
\label{eq:cat2}
\end{align}
where $\ket{L} = \ket{100}$, $\ket{C} = \ket{010}$ and $\ket{R} = \ket{001}$. The parent Hamiltonian of this state have $\ket{\varphi^1_\alpha} \pm \ket{\varphi^2_\alpha}$ as the two degenerate ground states.
This is in contrast with the eigenstates of Ref.~\onlinecite{Motrunich2019} ($\ket{\Phi_1}$ and $\ket{\Phi_2}$ in Sec.~\ref{sec:PXP}) which are injective MPSs, and thus unique ground states of their parent Hamiltonian.

 \subsubsection{Exact scars with $E=\pm \sqrt{q}$}
 
 We find that other (possibly degenerate) MPS scars appear at energies $E=\pm\sqrt{q}$ with $q$ integer. This property is a consequence of the structure of these matrix product states. Similarly to the case of periodic boundary conditions, the action of the Hamiltonian on these states is such that the complicated interaction is decoupled into smaller non-interacting blocks. Their energies are therefore determined by the energy of a single block: in the cases we consider, the energy of a block can be $0$ or $\pm \sqrt{q}$ where $q \le \alpha+1$ is the size of the block. In Appendix~\ref{sec:secondproof} we write down explicitly some exact eigenstates of $H_0^\alpha$ with energy $E=\pm \sqrt{2}$ for $\alpha=3$. 

\subsection{Relation with exact scars for $\alpha=1$}
The exact scars described here are reminiscent of the ones found in Ref.~\onlinecite{Motrunich2019}: there, it was shown that the PXP model ($\alpha=1$) has exact MPS scars at $E=0$ for periodic boundary conditions, and both at $E=0$ and $E=\pm \sqrt{2}$ when open boundary conditions are imposed. The states we study for $\alpha \ge 2$, however, show a qualitative difference with respect to them: in the case of open boundary conditions, the energy density profile does not have peaks at the edges, but has a pattern that is uniformly repeated in the full system. This can be understood from the MPS structure of these states. 
The scars in Eq.~\eqref{eq:scarsobc} have the form of AKLT states in which two-site blocks play the role of $S=1$ spin variables. As we show in Appendix~\ref{sec:edges}, the energy density of the PXP model corresponds to the local magnetization of the AKLT state in the $z$ direction. The boundary properties of the scars can be interpreted using the "dilute antiferromagnet" representation of the AKLT state: in the $S_z$ basis, the state is a superposition of configurations with alternating $+$ and $-$, and with an arbitrary number of $0$ placed in between. The different boundary vectors $\alpha,\beta$ of $\ket{\Gamma_{\alpha\beta}}$ fix the sign of the first and last non-zero spins of the configurations. Therefore, the local magnetization is non-zero close to the boundaries but goes to $0$ far from them.
The state in Eq.~\eqref{eq:s3}, on the other hand, has a very different structure: if we use, once again, a basis in which the local energy corresponds to a local magnetization, we can write $\ket{\psi_\alpha^{(3)}}$ as a superposition of configurations with a single $+$ (on one of the 3-site blocks), and $0$ magnetization everywhere else. Therefore, in contrast with the dilute antiferromagnet of the scars $\ket{\Gamma_{\alpha\beta}}$, this state is reminiscent of a spin wave, with a single magnetic excitation uniformly spread in the chain.

\subsection{Stability}
\label{subsec:fidsus}
We now analyse the response of the exact scars described above to a perturbation. The perturbation we apply is
\begin{multline}
    V^\alpha=\sum_i Z_{i-\alpha-1}P_{i-\alpha}\dots P_{i-1}X_iP_{i+1}\dots P_{i+\alpha}+\\
    P_{i-\alpha}\dots P_{i-1}X_iP_{i+1}\dots P_{i+\alpha}Z_{i+\alpha+1}.
    \label{eq:Valpha}
\end{multline}
This term has the same symmetries of $H_0^\alpha$, namely it commutes with $I$ and anticommutes with $C_{ph}$. Similarly to the PXP case, we use the fidelity susceptibility to check whether these states are stable to first order in perturbation theory.

\begin{figure}[h]
    \centering
    \includegraphics[width=0.45\textwidth]{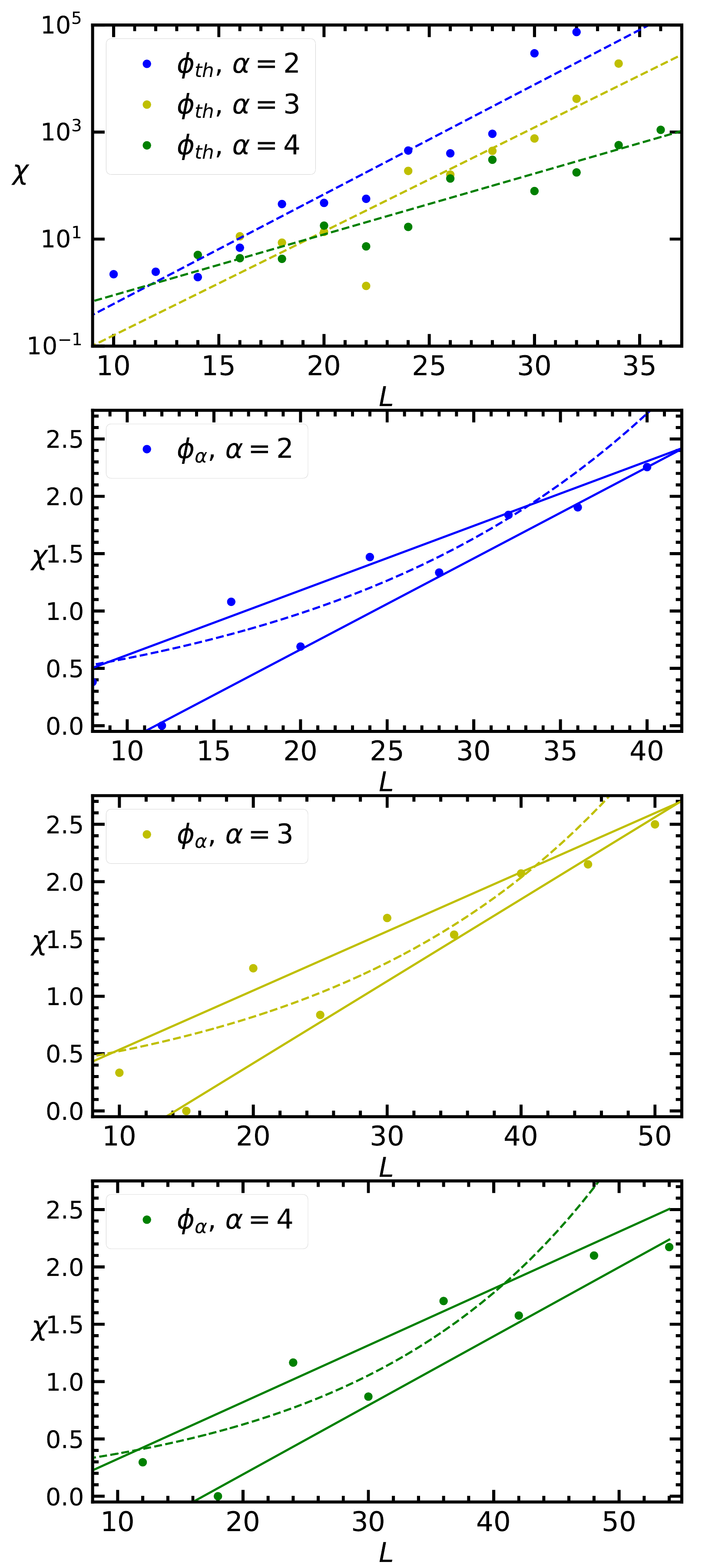}
    \caption{Scaling of the fidelity susceptibility with system size. The results shown refer to the generic states $\phi_{th}$ (upper panel) and the scarred eigenstates $\phi_\alpha$ (lower panels). Dashed lines are obtained from fits with an exponential scaling, solid lines with linear scaling. The result points at the same behavior occuring in the PXP model. }
    \label{fig:suscalpha}
\end{figure}

In Fig.~\ref{fig:suscalpha}, we present the results of the stability analysis. In the upper panel, we plot the fidelity susceptibility of a generic (thermal) eigenstate of the spectrum $\ket{\phi_{th}}$ (chosen as the eigenstate with energy closest to $1.9, 1.7, 1.35$ for $\alpha=2,3,4$ respectively): for every $\alpha$, the scaling with system size is exponential, as expected from ETH (dashed lines). In the lower panels, we plot instead the fidelity susceptibility of the scars $\ket{\phi_\alpha}$ defined in Eq.~\eqref{eq:scarphi}: the scaling here is linear\footnote{Similarly to the state $\ket{\Phi_{K=0}}$ in Fig.\ref{fig:fidsusc}-(d), the scaling for $\ket{\phi_\alpha}$ is subject to an even-odd effect related to the different parity under inversion symmetry ($I=(-1)^{L/(\alpha+2)}$).} (solid lines) for every $\alpha$, signalling a clear violation of ETH. These results suggest that the anomalous stability of the scars with $E=0$ is a generic feature of this class of one-dimensional models constrained by Rydberg blockade.

\section{Conclusions}
\label{sec:conclusions}
In this work, we investigated the stability against perturbations of exact quantum scars arising in spin chains constrained by Rydberg blockade. 
We first analysed the PXP model and found that some of the MPS scars found in Ref.~\onlinecite{Motrunich2019} exhibit a power law scaling of the fidelity susceptibility with system size. This result is a signature of their stability, a remarkable feature for eigenstates in the middle of a dense many-body spectrum. This fact is however limited to first order in perturbation theory, as a numerical analysis of the higher-order perturbative corrections reveals hybridization of exact scars eigenstates with thermal eigenstates.
{This behavior is reminiscent of the many-body "dark states" observed in Ref.~\onlinecite{Wurtz2020,Sugiura2020}.
We find the anomalous scaling of the fidelity susceptibility only }for scars with zero energy, suggesting that the properties of the $E=0$ subspace, such as the exponential degeneracy enforced by the invariance of this subspace under particle-hole and inversion symmetries, may be a key factor in stabilizing these states. Although not shown here, if we perturb with a term that breaks these properties, we find no signatures of stability for any of these low-entropy eigenstates.

To validate these conclusions, we extended our discussion to models with larger blockade radius $\alpha$. First, we constructed novel classes of states that are exact scars eigenstates for any $\alpha$ and have energy eigenvalues $E=0$ and $E=\pm\sqrt{q}$ (with $q$ integer). The construction is based on an effective decoupling of the sites of the chain into "non-interacting blocks", and allows us to write these states into simple matrix product form. We then studied their fidelity susceptibility under perturbations that do not spoil the exponential degeneracy of the zero-energy eigenspace, a common property of the family of constrained models we analysed. Also in this case, we found these eigenstates to be stable at first perturbative order when they belong to the $E=0$ subspace.

Our results suggest that an increasing number of exact MPS scars appear in the spectrum for larger values of $\alpha$, and their complete classification is beyond the scope of this work. It is also worth noticing that, contrarily to the $\alpha=1$ case (PXP model), no "approximate scars" eigenstates -- akin to the ones found in Ref.~\onlinecite{Turner2018} -- appear for $\alpha>1$, as can be seen from an inspection of the bipartite entanglement entropy of each eigenstate as a function of the energy. This fact provides strong indications that there is, in general, no relationship between the appearance of eigenstates with low entanglement entropy, equally spread uniformly in the energy spectrum, and the existence of exact MPS eigenstates in spin models constrained by Rydberg blockade. 
It stands as an open question whether these new exact MPS states can lead to clear experimental signatures, since, having no recurrent spectral structure, they are not expected to play any role in anomalous oscillations observed in experiments (that, indeed, were not reported for larger constraint radii). 

From a methodological standpoint, our results suggest that generalizations of the fidelity susceptibility to spectral properties can provide useful quantitative insights on the stability of ETH, in agreement to recent applications to quantum chaos diagnostics proposed in Ref.~\onlinecite{PhysRevX.10.041017,Polkovnikov2020}.

\begin{acknowledgements}
We acknowledge several useful discussions with G. Giudice, A. Polkovnikov, A. Scardicchio, P. Sierant, and J. Zakrzewski.
This work is partly supported by the ERC under grant number 758329 (AGEnTh), by the Quantera programme QTFLAG, and has received funding from the European Union's Horizon 2020 research and innovation programme under grant agreement No 817482 (Pasquans), and by the Italian Ministry of Education under the FARE programme MEPH. This work has been carried out within the activities of TQT. 

\end{acknowledgements}

\bibliographystyle{apsrev4-1}
\bibliography{bib}

\appendix

\section{Properties of the PXP and the other constrained models}
\label{sec:prop}
In this section, we summarize the properties of the spectrum of the PXP ($\alpha=1$) and the other constrained model with $\alpha>1$ of their pertubations.
For any $\alpha \ge 1$, the Hamiltonian $H_0^\alpha$ and the perturbation $V^\alpha$ commute with the space reflection symmetry $I$ and anticommute with the particle-hole symmetry $C_{ph}=\prod_i \sigma_i^z$.
This fact has some important consequences, that hold for any Hamiltonian with these symmetries:
 \begin{itemize}
     \item all the eigenstates with $E\neq 0$ are found in pairs of opposite energies ({\it doublets}), related by particle-hole symmetry ($C_{ph}\ket{E}=\ket{-E}$);
     \item states with $E=0$ can be classified as eigenstates of $C_{ph}$ ({\it singlets});
     \item the subspace of zero-energy eigenstates is exponentially large in $L$;
     \item the singlets have same eigenvalue with respect to $C_{ph}$ and $I$: this means that the zero-energy space is the direct sum of two subspaces with $C_{ph}=I=\pm 1$;
     \item if $\ket{\psi}$ and $\ket{\phi}$ are two singlet eigenstates of $H_0$, then $\braket{\phi|V|\psi}=0$. This holds even if $\braket{\phi|\psi}\neq 0$ (or even if $\ket{\psi}=\ket{\phi}$).
     
 \end{itemize}
 
 \subsection{Scars}
 Here we report the properties of the scars under the action of $I$ and $C_{ph}$.
 For the PXP model ($\alpha=1$), they satisfy:

\begin{align}
    &I\ket{\Gamma_{12}}=(-1)^{L/2-1} \ket{\Gamma_{12}}\\ &I\ket{\Gamma_{11}}=(-1)^{L/2} \ket{\Gamma_{11}} \\
    &C_{ph}\ket{\Gamma_{11}}=(-1)^{L/2} \ket{\Gamma_{11}} \\
    &I\ket{\Gamma_{21}}=(-1)^{L/2-1} \ket{\Gamma_{21}}\\ &I\ket{\Gamma_{22}}=(-1)^{L/2} \ket{\Gamma_{22}} \\
    &C_{ph}\ket{\Gamma_{22}}=(-1)^{L/2} \ket{\Gamma_{22}}.
\end{align}

The scars defined in Section \ref{subsubsec:sqrt3} for $\alpha>1$ and $L=(\alpha+2)n+3$ satisfy 
\begin{align}
    I\ket{\psi_\alpha^{(\pm 3)}}=(-1)^n \ket{\psi_\alpha^{(\pm 3)}}\\
    C_{ph}\ket{\psi_\alpha^{(\pm 3)}}=\ket{\psi_\alpha^{(\mp 3)}}.
\end{align}

\section{Stability to other perturbations}
\begin{figure}
    \centering
    \includegraphics[width=0.49\textwidth]{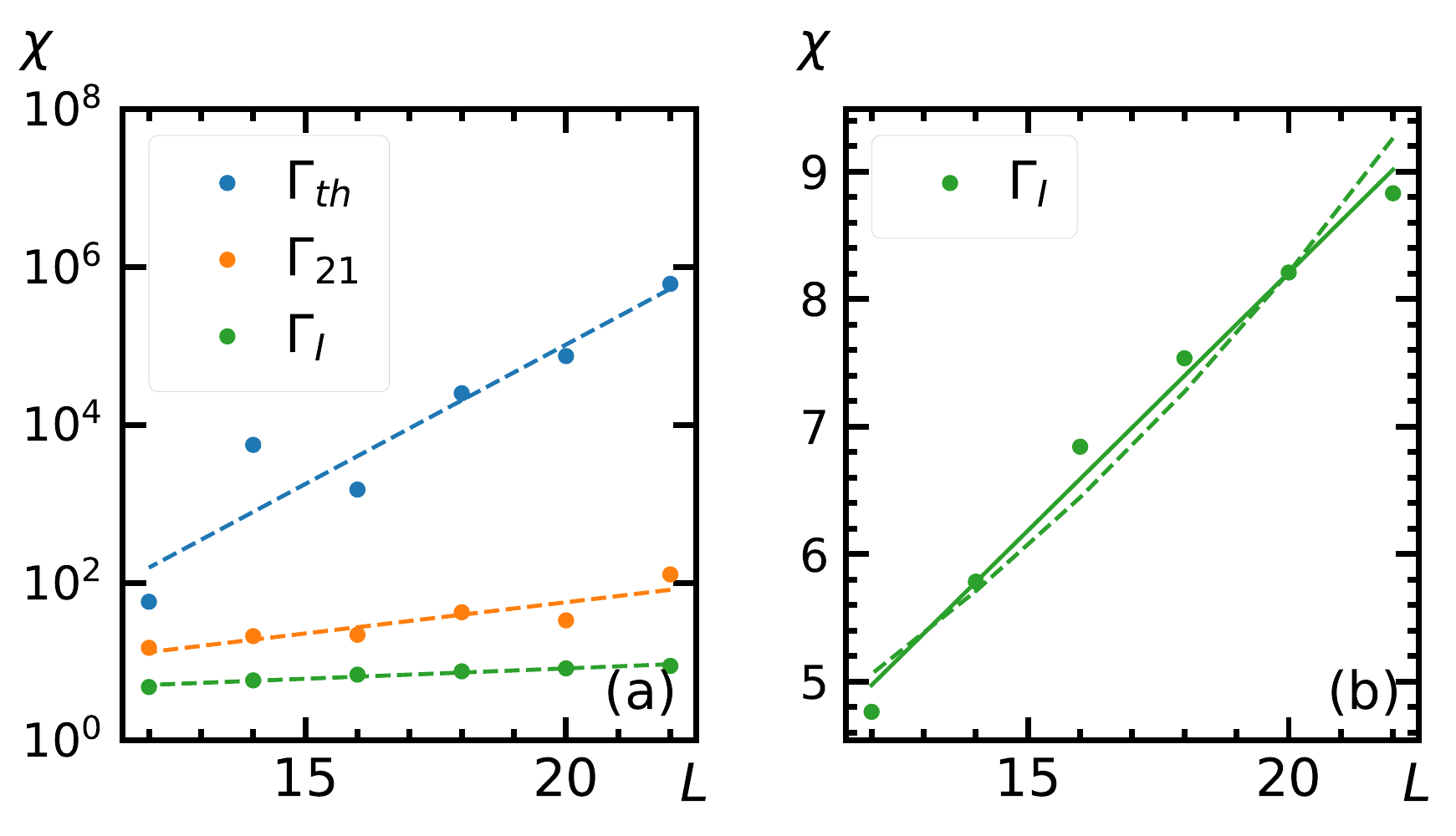}
    \caption{Scaling of the fidelity susceptibility with system size, for the perturbation $V'$. The results shown refer to the states (a) $\ket{\Gamma_{th}}$, $\ket{\Gamma_{21}}$ and (b) $\ket{\Gamma_{I}}$ with open boundary conditions. Dashed lines are obtained from fits with an exponential scaling, solid lines with linear scaling. Similarly to the results for the perturbation $V$ shown in Fig.~\ref{fig:fidsusc}, also in this case the scaling is exponential for the states $\ket{\Gamma_{th}}$, $\ket{\Gamma_{21}}$ (in agreement with ETH) and is linear for the state $\ket{\Gamma_{I}}$.}
    \label{fig:otherV}
\end{figure}

We report here the data of the fidelity susceptibility of the scars and of a generic thermal eigenstate in the PXP model for a different perturbation $V'$, defined as

\begin{equation}
    V'=\sum_{i=2}^{L-3} P_{i-2}\,\sigma^+_{i-1}\sigma^-_{i}\sigma^+_{i+1}P_{i+2}+\text{H.c.}
\end{equation}
The perturbation is again chosen in such a way to have the same properties under symmetry transformations as the PXP Hamiltonian $H_0$, i.e. $IV'I=V$, $C_{ph}V' C_{ph}=-V'$. The results in Fig.~\ref{fig:otherV} show the same behaviour that we observed for the perturbation $V$ in the main text: the fidelity susceptibility grows exponentially with system size for the states $\ket{\Gamma_{th}}$, $\ket{\Gamma_{21}}$ and linearly for the state $\ket{\Gamma_{th}}$.

\section{Exact scars in the PXP model -- properties of the edges}
\label{sec:edges}
In this section we recall some properties of the scars of Eq.~\eqref{eq:scarsobc} and \eqref{eq:scarspbc}, and we comment on the profile of the energy density. As was noticed in Ref.~\onlinecite{Motrunich2019}, the PXP Hamiltonian can be written as a sum of two parts: a part which contains two-body interactions between blocks, and one with single-block terms only. The two-body terms annihilate the scars (we refer to the appendix of Ref.~\onlinecite{Motrunich2019} for the proof), while the remaining terms are
\begin{equation}
    H'=\sum_b [\ket{10}\bra{00}+\ket{01}\bra{00}+h.c.]_b.
\end{equation}
A more convenient expression is obtained by defining the states
\begin{equation}
    \ket{\pm}=\frac{1}{2}(\ket{01}+\ket{10}+\sqrt{2}\ket{00}),
\end{equation}
\begin{equation}
    \ket{\mathbf{0}}=\frac{1}{\sqrt{2}}(\ket{10}-\ket{01}).
\end{equation}
The Hamiltonian $H'$ has the form
\begin{equation}
    H'=\sqrt{2} \sum_b (\ket{+}\bra{+}-\ket{-}\bra{-}).
\end{equation}
This expression is useful to interpret the profile of the energy density of the scars. After this change of basis and a gauge transformation with the unitary matrix $V=\frac{1}{\sqrt{2}}\begin{pmatrix}
1 & 1\\
1 & -1\end{pmatrix}$, the new matrices have the form
\begin{equation}
    A^+=V \frac{1}{2}(A^{01}+A^{10}+\sqrt{2}A^{00})V^{-1}=\begin{pmatrix}
    0 & \sqrt{2}\\
    0 & 0
    \end{pmatrix},
\end{equation}

\begin{equation}
    A^-=V \frac{1}{2}(A^{01}+A^{10}-\sqrt{2}A^{00})V^{-1}=\begin{pmatrix}
    0 & 0\\
    \sqrt{2} & 0
    \end{pmatrix},
\end{equation}

\begin{equation}
    A^{\mathbf{0}}=V \frac{1}{\sqrt 2}(A^{10}-A^{01})V^{-1}=\begin{pmatrix}
    1 & 0\\
    0 & 1
    \end{pmatrix},
\end{equation}

and the new boundary vectors are
\begin{equation}
    v_1'=Vv_1=\begin{pmatrix}
    1 \\ 0 \end{pmatrix},
\end{equation}

\begin{equation}
    v_2'=Vv_2=\begin{pmatrix}
    0 \\ 1 \end{pmatrix}.
\end{equation}

Now each block can be interpreted as a spin-1 variable with states $+,\mathbf{0}, -$ indicating the $S_z$ component, and the Hamiltonian $H'$ corresponds to the magnetization in the $z$ direction. The form of the matrices $A^+, A^-, A^{\mathbf 0}$, allows to easily see which are the non-zero components in the local $S_z$ basis: they are the ones with the structure of a "dilute antiferromagnet", i.e. with alternating $+$ and $-$ and an arbitrary number of $\mathbf{0}$s in between. This structure is a renowned feature of the AKLT state, whose relation with the MPS scars has been already pointed out in Ref.~\onlinecite{Motrunich2019}. In open boundary conditions, the boundary vectors fix the sign of the first non-zero spin: on the left $v_1'$ ($v_2'$) constrains it to be in a $+$ ($-$) state and viceversa for the vector on the right. Therefore, the components of the state $\Gamma_{12}$ have a number of $+$s that exceeds the number of $-$s by one, so its energy is $E=\sqrt{2}$ (and viceversa for $\Gamma_{21}$, with $E=-\sqrt{2}$). The states $\Gamma_{11}$ and $\Gamma_{22}$, on the other hand, have the same number of $-$s and $+$s, so they have energy $E=0$. The energy density profiles reported in Ref.~\onlinecite{Motrunich2019} can be understood as well from this construction: they correspond to the magnetization profile of the dilute antiferromagnet. In the bulk, the local magnetization averages to $0$, while on the boundary it is affected by the choice of the boundary vector.

\section{Exact scars with $E=\sqrt{3}$ -- Proof}
\label{sec:firstproof}
In this section we prove that the following state is an exact scar with energy $E=\sqrt{3}$ 
\begin{multline}
    \ket{\psi_\alpha^{(3)}}=\sum_{\vec s }\Big[ (1,0)^T \cdot N^{s_1}M^{s_2}\dots\\
    \dots M^{s_{2n}}N^{s_{2n+1}}\cdot (0,1) \Big] \ket{\vec s}
\end{multline}
where $s_1, s_2, \dots s_{2n+1}$ label the states of the blocks and

\begin{equation}
    M^s = \begin{cases}
    1 & \text{if } s=00\dots00\\
    0 & \text{otherwise},
    \end{cases}
\end{equation}

\begin{equation}
\label{eq:N1_2}
N^{0}=\begin{pmatrix}
0 & \sqrt{3}\\
0 & 0\end{pmatrix},
\qquad N^{L}=\begin{pmatrix}
0 & 1\\
0 & 1\end{pmatrix},
\end{equation}
\begin{equation}
\label{eq:N2_2}
N^{C}=\begin{pmatrix}
1 & 1\\
0 & -1\end{pmatrix},\qquad
N^{R}=\begin{pmatrix}
-1 & 1\\
0 & 0\end{pmatrix}.
\end{equation}

 The indices $0,L,C,R$ are the state of three-site block, with the following notation: $\ket{0}=\ket{000}$, $\ket{L}=\ket{100}$, $\ket{C}=\ket{010}$, $\ket{R}=\ket{001}$.

The matrices in Eqs.~\ref{eq:N1_2} and \ref{eq:N2_2} satisfy 
\begin{equation}
    N^{R} N^{L}=0,\qquad (N^{0}+N^{L}) (N^{R}+N^{0})=0.
\end{equation}
The first equation implies that the state satisfies the blockade constraint. We can split the Hamiltonian in two parts: $H=H_M+H_N$ where $H_M$ ($H_N$)
 flips only sites in the $M$ ($N$) blocks.
 
 We first prove that $H_M\ket{\psi_\alpha^{(3)}}=0$. Consider a single term $P_{i-\alpha}\dots P_{i-1}X_iP_{i+1}\dots P_{i+\alpha}$ where $i$ belongs to a block of type $M$: if $i$ is not the first or last site of the block, it can only be flipped if both neighbouring $N$ blocks are in the state $0$. However, this never happens because $N^0 M^s N^0=0$. If $i$ is the first site of the blocks, these two conditions must hold for it to be flippable: (i) the previous block must be in state $0$; (ii) the following block must be either in state $0$ or $R$. But $N^0 M^s N^0=N^0 M^s N^R=0$, so this Hamiltonian term annihilates the state. Similarly, using $N^0 M^s N^0=N^L M^s N^0=0$, we find that the last site of the block cannot be flipped. This means that the sites in the $M$ blocks are all "frozen" in the $0$ state and concludes the proof that $H_M\ket{\psi_\alpha^{(3)}}=0$.
 
 We now consider $H_N$:
 
\begin{multline}
    H_N\ket{\psi_\alpha^{(3)}}=\sum_{b}
    \Big[\big(\ket{0}\bra{R}\big)_b\big(1-\ket{L}\bra{L}\big)_{b+1}+\\
    \big(1-\ket{R}\bra{R}\big)_{b-1})\big(\ket{0}\bra{L}\big)_b +\\
    \big(\ket{0}\bra{R}\big)_b + h.c.\Big]\ket{\psi_\alpha^{(3)}}
\end{multline}
where $b=1,\dots n+1$ labels the blocks of type $N$. From the relations $N^R N^L= N^0 N^L + N^R N^0=0$, we find that all the terms involving more than one block cancel and we are left with

\begin{equation}
    H_N\ket{\psi_\alpha^{(3)}}=H' \ket{\psi_\alpha^{(3)}}.
\end{equation}
\begin{equation}
    H'=\sum_b \Big[\ket{0}\big(\bra{R}+\bra{C}+\bra{L}\big)+h.c.\Big]_b.
\end{equation}
Now, to prove that $H'\ket{\psi_\alpha^{(3)}}= \sqrt{3}\ket{\psi_\alpha^{(3)}}$, it is useful to change basis and define:
\begin{equation}
    \ket{\pm}=\frac{\ket{L}+\ket{C}+\ket{R}\pm\sqrt{3}\ket{0}}{\sqrt{6}},
\end{equation}
\begin{equation}
    \ket{l}=\frac{\ket{C}-\ket{L}}{\sqrt{2}}, \qquad
    \ket{r}=\frac{\ket{C}-\ket{R}}{\sqrt{2}}.
\end{equation}
In this new basis the matrices have the form

\begin{equation}
N^{+}=\begin{pmatrix}
0 & \sqrt{6}\\
0 & 0\end{pmatrix},
\qquad N^{-}=0,
\end{equation}
\begin{equation}
N^{l}=\begin{pmatrix}
1/\sqrt{2} & 0\\
0 & -\sqrt{2}\end{pmatrix},\;
N^{r}=\begin{pmatrix}
\sqrt{2} & 0\\
0 & -1/\sqrt{2} \end{pmatrix}.
\end{equation}
and the Hamiltonian $H'$
\begin{equation}
    H'=\sum_b \Big[\sqrt{3}\ket{+}\bra{+}-\sqrt{3}\ket{-}\bra{-}\Big]_b
\end{equation}
$H'$ is diagonal in the new basis $\{\ket{+}, \ket{-}, \ket{l}, \ket{r}\}$.
It is now sufficient to prove that all the non zero-components of $\ket{\psi}$ in the new basis have a one and only one block in $\ket{+}$ and all the others are in $\ket{l}$ or $\ket{r}$. This can be understood from the fact that (i) $N^+ N^{\alpha_1}\dots N^{\alpha_p} N^+=0$ (for any string in between) and that (ii) any string of matrices without $N^+$ is diagonal, so it annihilates when contracted with the boundary vectors $(1,0)^T$, $(0,1)$.
The energy density profile of this state is then easy to understand in these basis: all the three-site blocks have the same energy density, because the '$+$' can be located anywhere in the chain, while the other sites have energy density 0. This contrasts with the MPS scars found in Ref.~\onlinecite{Motrunich2019}: while there the energy density is localized on the edges because of the structure of dilute antiferromagnet, here the construction resembles a spin wave with a delocalized excitation.

\section{Exact scars with $E=\sqrt{2}$, $\alpha=3$}
\label{sec:secondproof}

\begin{figure}[h]
    \centering
    \includegraphics[width=0.49\textwidth]{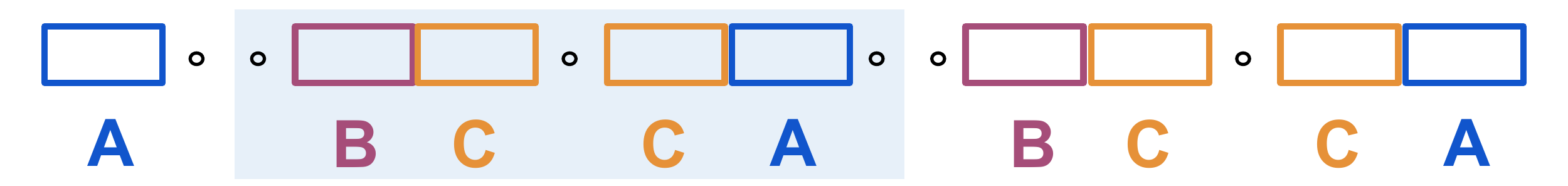}
    \caption{Structure of an MPS for $L=24$. The blocks are made of two sites. Empty dots are sites in the state $0$. The structure of the state for generic system sizes is based on the periodic repetition of the pattern $0BC0CA0$ (highlighted in the picture).}
    \label{fig:mps}
\end{figure}

We now consider the case $\alpha=3$ and construct exact eigenstates with $E=\pm\sqrt{2}$ as matrix product states with finite bond dimensions. They are constructed by assembling position dependent matrices in a periodic pattern, illustrated in Fig.~\ref{fig:mps}.

The matrices $A, B, C$ are defined on two-site blocks and have bond dimension 2. The dots represent empty sites. The pattern (0BC0CA0) that is repeated periodically consists of 11 sites. The first and last two sites of the open chain have to be in a block of type A or B. Therefore we have 4 possible states, labelled by the first and last block:
\begin{itemize}
    \item $\ket{\phi_{AB}^{(2)}}$, for $L=6+11n$;
    \item $\ket{\phi_{BA}^{(2)}}$, for $L=9+11n$;
    \item $\ket{\phi_{AA}^{(2)}}$ and $\ket{\phi_{BB}^{(2)}}$, for $L=13+11n$.
\end{itemize}

The matrices for the eigenvalue $E=\sqrt{2}$ are defined as

\begin{equation}
A^{00}=\begin{pmatrix}
    0 & 1/\sqrt{2}\\
    0 & 1
    \end{pmatrix}, \qquad
A^{10}=\begin{pmatrix}
    1/\sqrt{2} & 1/2\\
    0 & 0
    \end{pmatrix},
\end{equation}
\begin{equation}
A^{01}=\begin{pmatrix}
    -1/\sqrt{2} & 1/2\\
    0 & 0
    \end{pmatrix}
\end{equation}

\begin{equation}
B^{00}=\begin{pmatrix}
    1 & 1/\sqrt{2}\\
    0 & 0
    \end{pmatrix}, \qquad
B^{10}=\begin{pmatrix}
    0 & 1/2 \\
    0 & 1/\sqrt{2}
    \end{pmatrix},
\end{equation}
\begin{equation}
B^{01}=\begin{pmatrix}
    0 & 1/2 \\
    0 & -1/\sqrt{2}
    \end{pmatrix}
\end{equation}

\begin{equation}
C^{00}=\begin{pmatrix}
    0& 0\\
    1 & 0
    \end{pmatrix}, \qquad
C^{10}=\begin{pmatrix}
    0 & 1/\sqrt{2}\\
    0 & 0
    \end{pmatrix},
\end{equation}
\begin{equation}
C^{01}=\begin{pmatrix}
    0 & -1/\sqrt{2}\\
    0 & 0
    \end{pmatrix}
\end{equation}

The boundary vectors are obtained by contracting the extremal matrices with $(1,0)^T$ on the left and $(0,1)$ on the right. The states $\ket{\phi_{rs}^{(-2)}}=C_{ph} \ket{\phi_{rs}^{(2)}}$ ($r,s=A,B$) are other exact scars with energy $E=-\sqrt{2}$.

These scars satisfy the following properties:

\begin{align}
    I\ket{\phi_{AB}^{(\pm 2)}}=-\ket{\phi_{AB}^{(\pm 2)}}\\
    I\ket{\phi_{BA}^{(\pm 2)}}=-\ket{\phi_{BA}^{(\pm 2)}}\\
    I\ket{\phi_{AA}^{(\pm 2)}}=\ket{\phi_{BB}^{(\pm 2)}}\\
    I\ket{\phi_{BB}^{(\pm 2)}}=\ket{\phi_{AA}^{(\pm 2)}}.
\end{align}

\subsection{Proof}
We first prove that the state above satisfies the constraints. The conditions are: $B^{r}C^{s}=C^{r}A^{s}=0$ for $r=01,10$ and $s=01,10$, $C^{01}C^{01}=C^{01}C^{10}=C^{10}C^{10}=0$, and $A^{01}B^{10}=0$. It is straightforward to check that  all of them are satisfied by the matrices $A,B$ and $C$.

We now define the local Hamiltonian term $h_i=P_{i-3}P_{i-2}P_{i-1}X_i P_{i+1}P_{i+2}P_{i+3}$ and prove that $h_i\ket{\psi_{\alpha=3}}=0$ when $i$ is one of the sites between two $C$ blocks. To prove this, we note that $C^{00}C^{00}=0$, which immediately implies $P_{i-2}P_{i-1}P_{i+1}P_{i+2}\ket{\psi_{\alpha=3}}=0$. Similarly, we can prove that $h_i\ket{\psi_{\alpha=3}}=0$ when $i$ is one of the sites between an $A$ and a $B$ block by noting that $A^{00}B^{00}=0$ so the projectors in $h_i$ annihilate the state $\ket{\psi_{\alpha=3}}$. 

The next step is proving $h_i\ket{\psi_{\alpha=3}}=0$ for $i$ belonging to the $C$ blocks. To set the notation, we label the two-site blocks (of types $A$, $B$, $C$) in the chains with indices $b=0,1,2,\dots, N_b$ from left to right. We define $\Gamma_A$ as the set of integers $b$ such that the $b$-th block is of type $A$, and similarly for $\Gamma_B$ and $\Gamma_C$. We also define the operator $P_b^s$ which projects the block $b$ in the state $\ket{s}$. 

With this notation, we obtain the following equation
\begin{multline}
    \sum_{b\in \Gamma_C} \sum_{i \in b} h_i = \sum_{b,b+1\in \Gamma_C}P_{b-1}^{00} \ket{00}_{b}(\bra{10}+\bra{01})_{b}P^{00}_{b+1}\\
    +P_{b}^{00}\ket{00}_{b+1}(\bra{10}+\bra{01})_{b+1}P^{00}_{b+2}\\
    +P^{00}_{b-1} (\ket{10}+\ket{01})_{b}\bra{00}_{b}P^{00}_{b+1}\\
    +P^{00}_{b} (\ket{10}+\ket{01})_{b+1}\bra{00}_{b+1}P^{00}_{b+2}.
\end{multline}
The sum in the right hand side runs over the indices such that both $b$ and $b+1$ are blocks of type $C$. The first two terms of the sum annihilate $\ket{\psi_{\alpha=3}}$ because $C^{01}+C^{10}=0$, the last two terms because $C^{00}C^{00}=0$.

From the observations we made so far, we have now obtained that 
\begin{equation}
H\ket{\psi_{\alpha=3}}=\sum_{b\in \Gamma_A \cup \Gamma_B} \sum_{i \in b} h_i \ket{\psi_{\alpha=3}}.
\end{equation}
We can rewrite the action of these terms as
\begin{equation}
    \sum_{b\in \Gamma_A \cup \Gamma_B} \sum_{i \in b} h_i \ket{\psi_{\alpha=3}}= (H_{non-int}-H_{int})\ket{\psi_{\alpha=3}}.
\end{equation}
The Hamiltonian $H_{non-int}$ contains the terms
\begin{multline}
    H_{non-int}=\sum_{b\in \Gamma_A} P_{b-1}^{00}[\ket{00}(\bra{10}+\bra{01})+h.c.]_b\\
    + \sum_{b\in \Gamma_B} [\ket{00}(\bra{10}+\bra{01})+h.c.]_b P_{b+1}^{00},
\end{multline}
where, for the sake of brevity, in our notation for the boundary terms we choose to define $P_{-1}^{00}\equiv 1$, $P_{N_b+1}^{00}\equiv 1$.
The Hamiltonian $H_{int}$ reads
\begin{multline}
    H_{int}=\sum_{\substack{b \in \Gamma_A \\ b+1\in \Gamma_B}} P_{b-1}^{00}[\ket{00}\bra{01}+h.c.]_b P_{b+1}^{10}\\+
P_{b}^{01}[\ket{00}\bra{10}+h.c.]_{b+1}P_{b+2}^{00}.
\end{multline}
By noting that $A^{01}B^{10}=0$ and $C^{00}(A^{00}B^{10}+A^{01}B^{00})C^{00}=0$, we find that $H_{int}\ket{\psi_{\alpha=3}}=0$.

To conclude our proof, we now have to demonstrate that $H_{non-int}\ket{\psi_{\alpha=3}}=\sqrt{2}\ket{\psi_{\alpha=3}}$.
We define the states

\begin{equation}
    \ket{e}=\frac{\ket{10}+\ket{01}}{\sqrt{2}} \qquad \ket{o}=\frac{\ket{10}-\ket{01}}{\sqrt{2}},
\end{equation}

\begin{equation}
    \ket{\pm}=\frac{\ket{00}\pm\ket{e}}{\sqrt{2}},\qquad \ket{0}=\ket{00}.
\end{equation}

We now perform the following changes of basis:
on the $A$ and $B$ blocks, we use the (non-orthogonal) states $\ket{+}, \ket{o}, \ket{0}$, such that the new matrices of the MPS have the form

\begin{equation}
\tilde A^{+}=\begin{pmatrix}
    0 & 1\\
    0 & 0
    \end{pmatrix}, \qquad
\tilde A^{o}=\begin{pmatrix}
    1 & 0\\
    0 & 0
    \end{pmatrix},
\end{equation}
\begin{equation}
\tilde A^{0}=\begin{pmatrix}
    0 & 0\\
    0 & 1
    \end{pmatrix}
\end{equation}

\begin{equation}
\tilde B^{+}=\begin{pmatrix}
    0 & 1\\
    0 & 0
    \end{pmatrix}, \qquad
\tilde B^{o}=\begin{pmatrix}
    0 & 0\\
    0 & 1
    \end{pmatrix},
\end{equation}
\begin{equation}
\tilde B^{0}=\begin{pmatrix}
    1 & 0\\
    0 & 0
    \end{pmatrix},
\end{equation}

while on the $C$ blocks we use $\ket{0}$, $\ket{e}$ and $\ket{o}$, with the matrices

\begin{equation}
\tilde C^{0}=\begin{pmatrix}
    0 & 0\\
    1 & 0
    \end{pmatrix}, \qquad
\tilde C^e=0,\qquad
\tilde C^{o}=\begin{pmatrix}
    0 & 1\\
    0 & 0
    \end{pmatrix}.
\end{equation}

We now merge the pairs of consecutive $C$ blocks. The only non-zero matrices for the superblock are
\begin{equation}
\tilde G^{0,o}=\begin{pmatrix}
    0 & 0\\
    0 & 1
    \end{pmatrix}, \qquad
\tilde G^{o,0}=\begin{pmatrix}
    1 & 0\\
    0 & 0
    \end{pmatrix}.
\end{equation}

The components of $\ket{\psi_{\alpha=3}}$ now have the form
\begin{equation}
\label{eq:comp}
    \ket{\psi_{\alpha=3}}=\sum_{\vec s=(s_0,\dots, s_{N_b})}c_{\vec{s}}\ket{s_0}\otimes \ket{s_1}\dots\otimes \ket{s_{N_b}}
\end{equation}
where the sum runs over the three new states of the basis for each component $s_b$ and
\begin{equation}
\label{eq:cs}
    c_{\vec s}=
    \begin{pmatrix} 1 & 0 \end{pmatrix}
    \left(\dots \tilde B^{s_{b-1}}\tilde G^{s_{b}, s_{b+1}}\tilde A^{s_{b+2}}\tilde B^{s_{b+3}}\dots\right)
    \begin{pmatrix} 0 \\1\end{pmatrix}.
\end{equation}

From the simple structure of the matrices, it is now easy to see that the only cases that give $c_{\vec s}\neq 0$ are the ones where the product of matrices in parentheses is a sequence of $\tilde A^o$, $\tilde B^0$, $\tilde G^{o,0}$, followed by a single matrix $\tilde A^+$ or $\tilde B^+$ and then by a sequence of $\tilde A^0$, $\tilde B^o$, $\tilde G^{0,o}$. Consider now a state $\vec{s}$ that satisfies this condition and let $b^*$ be the index that corresponds to the $\tilde A^+$ or $\tilde B^+$ matrix. All the terms in $H_{non-int}$ annihilate $\ket{\vec{s}}$, except for the one with $b=b^*$: to prove this, it is sufficient to note that, for $b\in \Gamma_A$ if $(i)$ $b<b^*$ then $s_{b-1}=o$ and hence $P^0_{b-1}\ket{s_{b-1}}=0$, while if (ii) $b>b^*$ then $s_{b}=o$ and $[\ket{00}(\bra{10}+\bra{01})+h.c.]_b\ket{s_b}=0$; similarly, if (i) $b<b^*$ then $s_{b}=o$ and $[\ket{00}(\bra{10}+\bra{01})+h.c.]_b\ket{s_b}=0$, while if (ii) $b>b^*$ then $s_{b+1}=o$ and $P^0_{b+1}\ket{s_{b+1}}=0$.
The term of $H_{non-int}$ with $b=b^*$, on the other hand gives a non-zero term: if $b^* \in \Gamma_A$, then $s_{b^*-1}=0$ and $s_{b^*}=+$, so $P^0_{b^*-1}[\ket{00}(\bra{10}+\bra{01})+h.c.]_b\ket{\vec{s}}=\sqrt{2}\ket{\vec{s}}$, while if $b^* \in \Gamma_B$, then $s_{b^*+1}=0$ and $s_{b^*}=+$, so $[\ket{00}(\bra{10}+\bra{01})+h.c.]_b P^0_{b^*+1}\ket{\vec{s}}=\sqrt{2}\ket{\vec{s}}$. Therefore, we conclude that for each $\vec{s}$ such that $c_{\vec{s}}\neq 0$ $H_{non-int}\ket{\vec{s}}=\sqrt{2}\ket{\vec{s}}$, and using Eq.~(\ref{eq:comp}), we have $H_{non-int}\ket{\psi_{\alpha=3}}=\sqrt{2}{\psi_{\alpha=3}}$.

\end{document}